\documentclass{article}
\usepackage{amsmath}

\input{tcilatex}

\begin{document}

\textbf{VACUUM RESPONSE TO COSMIC STRETCHING: ACCELERATED UNIVERSE AND
PREVENTION OF SINGULARITY}

\bigskip

\bigskip

E. A. Novikov

\bigskip

Institute for Nonlinear Science, University of California - San Diego, La
Jolla, CA 92093 - 0402

\bigskip

Spacetime stretching is included in the general relativity alongside with
the spacetime curvature. Response of the vacuum to cosmic stretching is
considered as macroscopic quantum effect. This effect explains the
accelerated expansion of the Universe without resorting to Plank scale. For
negative stretching (collapse) the same effect can prevent formation of
singularity. Stretching effect can be important for a variety of cosmic
phenomena, including collisions of galaxies and local collapses.

\bigskip

\bigskip

The Einstein equation of the general relativity (1) postulates a balance
between the spacetime curvature and the energy-momemtum. Initially, Einstein
introduced the cosmological constant (CC) $\Lambda $ into the equation to
make the steady state cosmology. When the expansion of the Universe was
discovered, Einstein dismissed CC as his greatest blunder. Recent
observations of the type Ia supernovae [1,2] suggest the accelerated
expansion of the Universe (AU), which renewed interest to CC. However, there
is a problem with $\Lambda $, which has dimension $L^{-2}$. In the classical
theory we can not construct a scale from the gravitational constant $G$ and
the velocity of light $c$. In the quantum theory the Plank constant 
h{\hskip-.2em}\llap{\protect\rule[1.1ex]{.325em}{.1ex}}{\hskip.2em}
leads to the Plank scale $L_{P}=(\frac{G\text{%
h{\hskip-.2em}\llap{\protect\rule[1.1ex]{.325em}{.1ex}}{\hskip.2em}%
}}{c^{3}})^{\frac{1}{2}}$. However, the cosmological observations suggest $%
\Lambda L_{P}^{2}\approx 10^{-123}$(see reviews in Refs. 3 and 4).

In this Letter we introduce spacetime stretching into the general relativity
alongside with the curvature. Such modification of the theory produces a
natural dynamical scaling for AU and, potentially, eliminates the problem
with CC. It also gives a new role to the vacuum in cosmic phenomena.
Particularly, stretching effect can prevent formation of singularity.

The Einstein equations for the general relativity have the form [5,6]:

\begin{equation}
R_{ik}-\frac{1}{2}Rg_{ik}+\Lambda g_{ik}=-\frac{8\pi G}{c^{4}}T_{ik}  \tag{1}
\end{equation}

Here $R_{ik}$ is the Ricci tensor, $R$ is the scalar curvature, $g_{ik}$ is
the spacetime metric tensor and $T_{ik}$ is the energy-momentum tensor. This
famous and beautiful equation serves very well in cosmology. But the problem
with AU and $\Lambda $ suggests that something is missing in equation (1). $%
\Lambda $ represents influence of the vacuum. The traditional assumption
that $\Lambda $ is constant means that on macroscopic level the vacuum is
irresponsive to what is going on in the Universe. Perhaps, this is incorrect
and $\Lambda $ is not a constant, but a function of some dynamical
variables. Below we will present our choice of such variables.

The singular vortices, sources (sinks) and vortex-sinks are well known
models in fluid dynamics (including geophysical fluid dynamics), magnetized
plasma, superfluidity and superconductivity (see, for example, Refs. 7 - 9
and references therein). Dynamics of distributed vortices is a well
developed area of research. However, to our knowledge, dynamics of
distributed sources (DS) has not been considered until recently [10 -12].
Local intensity of nonrelativistic DS is characterized by the divergency of
the 3-velocity field $\partial v^{\alpha }/\partial x^{\alpha }$ ( summation
over the repeated Greek indexes is assumed from 1 to 3 ). Dynamical equation
for DS was obtained by considering superposition of localized sources, which
move each other with induced velocity field [8 - 10]. One of the
applications of DS is cosmology and, in particular, the problem of AU. The
relativistic equation for DS can be written in the form [11]:

\begin{equation}
\frac{d\sigma }{ds}=f,\;\sigma =u_{;\;i}^{i}\equiv \frac{\partial u^{i}}{%
\partial x^{i}}+\Gamma _{ki}^{i}u^{k},\;\frac{d}{ds}\equiv u^{k}\frac{%
\partial }{\partial x^{k}}  \tag{2}
\end{equation}

Here $u^{i}$ is the 4-velocity field, position of fluid element is
characterized by the 4-vector $x^{i}$ with components ($\tau =ct,x^{\alpha }$%
), $\sigma $ is the covariant divergency of the 4-velocity field, summation
over repeated Latin indexes is assumed from $0$ to $3$. The Christoffel
symbols satisfy condition [5,6]:

\begin{equation}
\Gamma _{ki}^{i}=\frac{1}{2g}\frac{\partial g}{\partial x^{k}}  \tag{3}
\end{equation}%
where $g$ is the determinant of the metric tensor. The forcing $f$ \ may
include diffusion and some other effects, particularly, with $f\sim \sigma
^{2}$ (see Ref. 10 and below).

The case of "free" DS with $f=0$ was considered in Refs. 10 and 11. It was
shown that the effect of DS, described by equation (2), is similar to the
effect of CC in the general relativity ( see also below). It seems natural
to assume that relativistic DS are produced by the quantum vacuum [11].
However, in macroscopic description of the Universe we have dynamical scales
much larger than the scale $L_{P}$.

The divergency $\sigma $ is an important dynamical variable, which
characterizes the spacetime stretching of the media. It is the only
dynamical characteristic of the media, which enters into the balance of the
proper number density of particles $n$ : $dn/ds+\sigma n=q$, where $q$ is
the rate of particle production (or absorption).

\textbf{Now, suppose you are the vacuum.} Your major microscopic activity is
to produce and absorb particles. At the same time, you would like to
prevent, say, macroscopic breakdown of spacetime ( it could be broken at
scale $L_{P}$). Your macroscopic representative in equation (1) is scalar $%
\Lambda $. It seems natural, that you will make $\Lambda $ a function of
dynamical scalars, which are involved in the production (absorption) of
particles. Stretching ($\sigma $) is such scalar and, perhaps, its tendency
also can be useful.

Having in mind that $\sigma $ has dimension $L^{-1}$ and $d\sigma /ds$ has
dimension $L^{-2}$, in simplest approximation, we get:

\begin{equation}
\Lambda =\beta \frac{d\sigma }{ds}+\gamma \sigma ^{2}+\lambda  \tag{4}
\end{equation}

Here $\beta $ and $\gamma $ are nondimensional constants (presumably, of the
order of $1$), constant $\lambda $ has dimension $L^{-2}$. Eventually, we
would like to put $\lambda =0$. Insertion of (4) into (1) gives linear
superposition of equations (1) and (2) with $f\sim \sigma ^{2}$. The first
two terms in expression (4) can produce a dynamical scaling for the problem
of AU. They can compete with the curvature (see below) and they are asking
to be included into the equation (1). There is a similarity between
expression (4) and the left hand side of equation (1). Stretching in (4)
play a role of curvature in (1). This similarity becomes even more
transparent in application to problems of AU and collapse (see below
equations (8) and (9)).

It is tempting to speculate about the nature of the vacuum, based on the
expression (4). From everyday experience one can relate stretching to a
release of energy (see below equation (15)). Perhaps, this analogy can help
to develop a theory of the vacuum and to calculate $\beta $ and $\gamma $
from first principles. What we have in mind is a sort of superfluid with
maxima/minima of DS (stretching) as leading centers. Near these centers we
can expect production/absorption of unknown particles.

For the global quantum description of the Universe we need special tools.
One promising possibility is the topos approach (see recent paper [13] and
references therein). Having model (4) can be useful in this or similar
approaches. It is, of course, a long perspective ( see also discussion after
formula (16)). In a mean time, we can consider ($\beta ,\gamma $) as
semiempirical constants and look for cosmic phenomena, where stretching
could be important. In what follows, we will obtain some solutions and put
certain limitations on $\beta $ and $\gamma $, based on physical
requirements and cosmic observations.

For homogeneous isotropic AU in the proper synchronized frame of reference
[5,6] we have: $u^{0}=1,u^{\alpha }=0$. From (2) and (3) we get:%
\begin{equation}
\sigma =\frac{1}{2g}\frac{\partial g}{\partial \tau }=\frac{3}{a}\frac{%
\partial a}{\partial \tau }  \tag{5}
\end{equation}%
where $a(\tau )$ is the scale factor and we took into account that $g\sim
a^{6}$.

Inserting (4) into (1) and using standard procedure [5,6], we get two
equations:

\begin{equation}
2\frac{a^{\prime \prime }}{a}+(\frac{a^{\prime }}{a})^{2}+\frac{k}{a^{2}}%
-\lambda -\beta \sigma ^{\prime }-\gamma \sigma ^{2}=-\frac{8\pi G}{c^{4}}p 
\tag{6}
\end{equation}

\begin{equation}
(\frac{a^{\prime }}{a})^{2}+\frac{k}{a^{2}}-\frac{1}{3}(\lambda +\beta
\sigma ^{\prime }+\gamma \sigma ^{2})=\frac{8\pi G}{3c^{4}}\varepsilon 
\tag{7}
\end{equation}

Here prime indicates differentiation over $\tau $, $k=0,\pm 1$ is the
discrete curvature parameter, $p$ is the pressure and $\varepsilon $ is the
energy density. Using (5), we rewrite equations (6) and (7) in the form:

\begin{equation}
(2-3\beta )\frac{a^{\prime \prime }}{a}+(1+3\beta -9\gamma )(\frac{a^{\prime
}}{a})^{2}+\frac{k}{a^{2}}-\lambda =-\frac{8\pi G}{c^{4}}p  \tag{8}
\end{equation}

\begin{equation}
-\beta \frac{a^{\prime \prime }}{a}+(1+\beta -3\gamma )(\frac{a^{\prime }}{a}%
)^{2}+\frac{k}{a^{2}}-\frac{\lambda }{3}=\frac{8\pi G}{3c^{4}}\varepsilon 
\tag{9}
\end{equation}

To these equations we need to add an equation of state. We start with the
simplest equation of state $p=0$ (dust approximation), which can be used for
the present epoch. From (8) we will obtain $a(\tau )$ and then insert it
into (9) to get $\varepsilon (\tau )$. We choose $k=0$, which corresponds to
the observations. Let us try to get away without $\lambda $. So, we put $%
\lambda =0$ in order to avoid dimensional scaling and corresponding
ridiculously small numbers (see above).

First of all, we have special case: $\beta =2/3,\;(1-3\gamma )(a^{\prime
}/a)^{2}=0$. If $\gamma \neq 1/3$, we got stationary Universe. In what
follows we assume $\beta \neq 2/3$.

From (8) we have positive acceleration ( $a^{\prime \prime }>0$), if:

\begin{equation}
\theta \equiv \frac{3(1-3\gamma )}{2-3\beta }<1  \tag{10}
\end{equation}

Using variable $H=a^{\prime }/a$, we get from (8):

\begin{equation}
H^{\prime }+\theta H^{2}=0,\;H(\tau )=\frac{H_{0}}{1+H_{0}\theta \tau } 
\tag{11}
\end{equation}

It is nice that two constants $\beta $ and $\gamma $ are combined into only
one constant $\theta $. Integration of (11) gives:

\begin{equation}
a(\tau )=a_{0}(1+H_{0}\theta \tau )^{1/\theta }  \tag{12}
\end{equation}

From (9) we now get:

\begin{equation}
\frac{4\pi G}{c^{4}}\varepsilon (\tau )=\theta H^{2}(\tau )  \tag{13}
\end{equation}

Positive $\varepsilon $ leads to $\theta >0$. In combination with (10), we
have:

\begin{equation}
0<\theta <1  \tag{14}
\end{equation}

From observations it is known that $H_{0}$ $>0$. From (12) we got the
finite-time singularity in the past ($\tau =-(H_{0}\theta )^{-1}$). However,
the dust approximation is not valid near singularity (see below). We can use
(12) with condition (10) only for the time period when there was
acceleration (see below). For the future from (12) we have the power law
accelerated expansion. The same result (12) we get from equation (2) with
the effect of gravitation, represented by forcing $f=-\frac{\theta }{3}%
\sigma ^{2}$. "Free" DS with $f=0$\ \ \ produce exponential expansion
[10,11].\ 

Let us estimate the total energy of the Universe by using equations (11) -
(13):

\begin{equation}
E(\tau )\sim \varepsilon (\tau )a^{3}(\tau )\sim \frac{\theta
c^{4}a_{0}^{3}H_{0}^{2}}{G}(1+H_{0}\theta \tau )^{\frac{3}{\theta }-2} 
\tag{15}
\end{equation}

Taking into account (14), we see that the total energy is increasing. This
supply of energy is due to the stretching effect, because with $\beta
=\gamma =0$ we have $\theta =3/2$ and $E(\tau )=E_{0}$. Energy is conserved
( $\theta =3/2$ ) in more general case when $\beta =2\gamma $. From (5) and
(12) it follows that in this case the stretching terms compensate each other
in expression (4).

Multiplication (11) and (12), or differentiation of (12), gives:

\begin{equation}
a^{\prime }(\tau )=a_{0}H_{0}(1+H_{0}\theta \tau )^{1/\theta -1}  \tag{16}
\end{equation}

From condition (14) it follows that asymptotically the expansion becomes
superluminary (compare with Ref. 11). Superluminary propagation is usually
associated with imaginary $\varepsilon $ and corresponding tachyon field.
Such features can be described with the model (4), if constants have
imaginary components (this work is in progress).

\textbf{\{ }Note, that imaginary fields (IF) have been used recently [14] to
eliminate classical electromagnetic divergencies, namely, the infinite
self-energy of electrons and the paradoxical self-acceleration. The same
(algebraic) approach works for the quantum interaction of charges. The
natural, but more difficult, next step is to eliminate infinities in quantum
field theories with an appropriate use of IF.

Another application of IF is to the phenomena of consciousness ( considered
as collective effect of billions of interconnected nonlinear neurons). The
brain activity revealed the regime of scale-similarity [15-17], which is
typical for systems with strong interaction of many degrees of freedom (
particularly, for turbulence [18]). Modeling of the effects of consciousness
on the electric currents in the human brain leads to the use of IF [19].
Possible connection of DS ( stretching ) with consciousness was indicated in
Ref. 19. Now we have additional reason (16) in favor of such connection.
Another possible connection is between IF, which eliminate electromagnetic
divergencies [14], and IF in the modeling of consciousness [19].

Finally, consider the quantum entanglement, the EPR experiment and all that
( see corresponding discussions in Refs. 20, 21 ). According to (4) and
(16), the vacuum - stretching interaction leads to the superluminary IF -
effect on cosmological scale. Why the same vacuum can not produce an IF -
signal during the process of quantum measurement? This will explain the
quantum entanglement. Perhaps, the macroscopic quantum effect (4) and IF
provide the missing link between the quantum theory and the general
relativity.\textbf{\}}

Cosmic observation suggests that in the past there was deceleration. We can
explain this in different ways. First of all, the expression (10) for the
parameter $\theta $ was obtained in the dust approximation ( $p=0$ ). Let us
consider the ultra-relativistic equation of state $p=\varepsilon /3$. From
(8) and (9), \ with $\lambda =0$ and $k=0$, we get solution similar to (11)
with $\theta $ replaced by

\begin{equation}
\theta _{\ast }=\frac{2(1-3\gamma )}{1-2\beta }  \tag{17}
\end{equation}

For the energy we now get expression (13) with $\theta $ replaced by $\theta
_{\ast }/2$. We can satisfy conditions (14) and $\theta _{\ast }>1$ with the
same parameters $\beta $ and $\gamma $, if: $\beta +1/3<3\gamma <\beta +1/2<1
$. Thus, we can get acceleration in the present epoch and in the future
(with $p=0$) and deceleration in the past (with $p=\varepsilon /3$). A more
complicated analysis for intermediate equations of state and fitting the
data are beyond the scope of this Letter.

So far, we assumed that the stretching coefficients $\beta $ and $\gamma $
are universal - the same for all epochs and for all equations of state. Any
softening of this strong assumption will make it easier to fit the data. For
description of the accelerated Universe, there is a choice: play with $\beta 
$ and $\gamma $ (of the order of $1$) or restore $\lambda $ (small numbers).
Of course, it is not only the problem of fitting the data with reasonable
coefficients. Equation (4) suggests a new physics with a more active role of
the vacuum in cosmic phenomena.

Particularly, consider closed model ($k=1$) with $p=0$. From equation (8)
with $\lambda =0$ we have:

\begin{equation}
(2-3\beta )aa^{\prime \prime }+(1+3\beta -9\gamma )(a^{\prime })^{2}+1=0 
\tag{18}
\end{equation}%
The first integral of this equation can be written in the form:

\begin{equation}
(a^{\prime })^{2}+\mu =[(a_{0}^{\prime })^{2}+\mu ](\frac{a}{a_{0}}%
)^{2(1-\theta )}  \tag{19}
\end{equation}%
where $\mu =(1+3\beta -9\gamma )^{-1}$. Note, that without stretching effect
($\beta =\gamma =0$) we have $\mu =1$, $\theta =3/2$. With such parameters,
gravitational collapse ($a_{0}^{\prime }<0$), according to (19), will lead
to singularity ($a\rightarrow 0$, $a^{\prime }\rightarrow -\infty $).
Stretching effect can prevent singularity. We assume, as above, that $\theta
<1$ and $\mu >0$. Then, from (19) it follows that gravitational collapse ($%
a_{0}^{\prime }<0$) will lead to a finite core with:

\begin{equation}
a_{\ast }=a_{0}\left[ \frac{\mu }{(a_{0}^{\prime })^{2}+\mu }\right] ^{\frac{%
1}{2(1-\theta )}}  \tag{20}
\end{equation}

We presented the simplest description of the dynamics of the Universe with
introduction of spacetime stretching into the Einstein equations. More
general cases can be studied, based on equations (8) and (9). Even more
general cases can be considered by using (1) and (4). Stretching effect can
play an important role not only in the global description of the Universe,
but also in a variety of cosmic phenomena, including collisions of galaxies
and local collapses.

\bigskip

REFERENCES

\bigskip

[1] A. G. Riess et al., Astron. J.\textbf{\ 116}, 1009 (1998); Astrophys.
J., \textbf{607}, 665 (2004)

[2] S. Perlmutter et al., Astrophys. J. \textbf{517}, 565 (1999)

[3] T. Padmanabhan, Phys. Reports \textbf{380}, 235 (2003)

[4] D. H. Cole, Class. Quantum Gravity \textbf{22}, R125 (2005)

[5] L. D. Landau and E. M. Lifshitz, The Classical Theory of Fields,
Pergamon Press, 1987

[6] J. V. Narlikar, An Introduction to Cosmology, Cambridge University
Press, 2002

[7] E. A. Novikov, Ann. N. Y. Acad. Sci. \textbf{357}, 47 (1980)

[8] E. A. Novikov and Yu. B. Sedov, Fluid Dyn. \textbf{18}, 6 (1983)

[9] A. E. Novikov and E. A. Novikov, Phys. Rev. E \textbf{54}, 3681 (1996)

[10] E. A. Novikov, Physics of Fluids \textbf{15}, L65 (2003)

[11] E.\ A. Novikov, arXiv:nlin.PS/0511040

[12] E. A. Novikov, arXiv:nlin.PS/0608050

[13] C. J. Isham, arXiv:quant-ph/0508225

[14] E. A. Novikov, arXiv:nlin.PS/0509029

[15] E. Novikov, A. Novikov, D. Shannahof-Khalsa, B. Schwartz, and J.
Wright, Phys Rev. E \textbf{56}(3), R2387 (1997)

[16] E. Novikov, A. Novikov, D. Shannahof-Khalsa, B. Schwartz, and J.
Wright, in Appl. Nonl. Dyn. \& Stoch. Systems (Eds. J. Kadtke \& A.
Bulsara), p. 299, Amer. Inst. Phys., N. Y., 1997

[17] W. J. Freeman, L. J. Rogers, M. D. Holms, D. L. Silbergelt, J.
Neurosci. Meth. \textbf{95}, 111 (2000).

[18] E. A. Novikov, Phys. Rev. E \textbf{50}(5), R3303 (1994)

[19] E. A. Novikov, arXiv:nlin.PS/0309043; arXiv:nlin.PS/0311049;
arXiv:nlin.PS/0403054; Chaos, Solitons \& Fractals, \textbf{25}, 1 (2005);
arXiv:nlin.PS/0502028

[20] D. Bohm \& B. J. Hiley, The Undivided Uviverse, Routledge 1993

[21] Roger Penrose, The Road to Reality, Jonathan Cape 2004

\end{document}